\documentclass[10pt]{iopart}
\usepackage{times}
\usepackage{iopams}
\usepackage{mathptmx}
\usepackage[dvips]{graphicx}
\newcommand{\ped}[1]{\ensuremath{_{\rm #1}}}
\begin{document}

\title[Point-Contact Spectroscopy in RuSr$_2$GdCu$_2$O$_8$]
{Point-contact study of gap amplitude and symmetry in
RuSr$_2$GdCu$_2$O$_8$}

\author{G.\@A.
Ummarino~\dag\footnote[6]{To whom correspondence should be
addressed (giovanni.ummarino@infm.polito.it)}, A. Calzolari~\dag,
D. Daghero~\dag, R.\@S. Gonnelli~\dag, V.\@A. Stepanov~\ddag,
R.\@Masini~\S, M.\@R. Cimberle~$\|$ }

\address{\dag INFM -- Dipartimento di Fisica, Politecnico di Torino,
10129 Torino, Italy}

\address{\ddag P.N. Lebedev Physical Institute, Russian Academy of
Sciences, 119991 Moscow, Russia}

\address{\S CNR - IENI, Sezione di Milano, 20125, Milano, Italy}

\address{$\|$ IMEM-CNR Sezione di Genova, 16146 Genova, Italy}

\begin{abstract}
We present the results of point-contact spectroscopy measurements
in polycrystalline RuSr$_2$GdCu$_2$O$_8$ samples previously
characterized from both the structural and the electrical point of
view. AC susceptibility and resistivity measurements, as well as
SEM and optical microscopy, indicate that the polycrystalline
material is made up of small, weakly connected grains. Possibly
because of this morphology, point-contact measurements turned out
to be rather difficult. Best reproducibility of the conductance
curves was achieved in point contacts obtained by pressing Au or
Pt-Ir tips on the surface of the samples. The low-temperature
conductance curves reproducibly show a clear zero-bias cusp that
might be interpreted as due to Andreev reflection at the normal
metal-superconductor interface in the case where the
superconducting order parameter has a pure $d$-wave symmetry -
even though alternative explanations cannot be ruled out. In this
paper we show that these curves can indeed be fitted, in the
framework of the generalised BTK model, by using pure $d$-wave
symmetry and a gap of the order of 6 meV. Finally, we present the
temperature dependence of the conductance curves from 4.2 K up to
the critical temperature of the junctions.
\end{abstract}
\section{Introduction}
The hybrid ruthenate-cuprate superconductor
RuSr$\ped{2}$GdCu$\ped{2}$O$\ped{8}$ (Ru-1212) is a triple
perovskite material comprising CuO$\ped{2}$ bilayers and
RuO$\ped{2}$ monolayers. It was originally synthesized
\cite{Bauernfeind} with the purpose of incorporating a metallic
layer between the CuO$_2$ planes, so as to enhance the inter-layer
coupling and reduce the anisotropy.  The role of the
ruthenium-oxide layers was thus intended to be similar to that of
the one-dimensional Cu-O chains in YBCO. In these layers, the
ruthenium atoms present the same square-planar coordination as in
the CuO$_2$ planes, with a similar bond length. As evidence by
X-ray absorption near edge structure (XANES) \cite{Liu}, nuclear
magnetic resonance (NMR) \cite{Kumagai,Tokunaga} and magnetization
measurements \cite{Butera} ruthenium occurs in a mixed valence
state as Ru$^{4+}$ and Ru$^{5+}$ with concentrations equal to 40\%
and 60\% respectively. As a result, RuO$_2$ planes also act as a
charge reservoir, doping the superconducting CuO$_2$ planes.

In addition to the superconducting transition, the compound
features magnetic order below the temperature $T\ped{M}\simeq
133$~K, whose nature is still controversial. After the first
results suggesting a ferromagnetic order \cite{Bernhard,McCrone}
some theoretical and experimental arguments have been reported in
favour of a antiferromagnetic ordering \cite{Lynn,Nakamura}. One
possibility to reconcile these opposite evidences is to admit that
the ferromagnetically-ordered RuO$_2$ planes are coupled to one
another via an antiferromagnetic interaction \cite{Butera}.

Whatever its nature, the magnetic order is found to survive even
when superconductivity sets in. While the coexistence of
antiferromagnetism and superconductivity is common among cuprates,
the coexistence of superconductivity and ferromagnetism would
contrast the widespread belief that these two ordering are
competing and mutually exclusive. This is the more true if one
takes into account that, according to muon-spin rotation ($\mu
SR$) measurements \cite{Bernhard}, the ferromagnetic order looks
uniform and homogeneous, even below $T\ped{c}$, on a scale of
typically 20\AA. This would imply that the interaction between the
superconducting and the ferromagnetic order parameter is very
weak. As a matter of fact, high-temperature susceptibility data
indicate that the ferromagnetic order involves only the Ru
moments~\cite{Bernhard} while Zn substitutions suggest that only
the CuO$_2$ planes host the superconducting charge
carriers~\cite{Bernhard}. The decoupling of CuO$_2$ and RuO$_2$
layers has been recently confirmed by a comprehensive study of DC
and AC susceptibility, DC resistance, magnetoresistance , Hall
effect and microwave absorption \cite{Pozek}, showing that the
RuO$_2$ planes are conducting but do not develop
superconductivity.

In spite of the great experimental and theoretical efforts focused
on the interplay between superconductivity and magnetism, to our
knowledge no spectroscopic studies of the superconducting order
parameter in  RuSr$\ped{2}$GdCu$\ped{2}$O$\ped{8}$ have been
reported so far in literature. In this paper we present and
discuss the results of point-contact spectroscopy measurements in
polycrystalline RuSr$_2$GdCu$_2$O$_8$ previously characterized by
SEM and optical microscopy, AC susceptibility and resistivity
measurements. We will show that the order-parameter symmetry that
best fits the experimental conductance curves is $d$-wave, thus
indicating that, as in the other cuprates, superconductivity in
Ru-1212 is due to singlet Cooper pairs.

\section{Experimental details}
\subsection{Preparation and characterization of the samples}
The single-phase, polycrystalline samples we used for our
measurements were synthesized by solid state reaction starting
from high purity stoichiometric powders of RuO$\ped{2}$,
Gd$\ped{2}$O$\ped{3}$, CuO and SrCO$\ped{3}$. The raw materials
were reacted in air at about 960 $^\circ$C, in order to decompose
the SrCO$\ped{3}$. Then, they were heated in N$\ped{2}$ flow at
1010 $^\circ$C and annealed in O$\ped{2}$ flow at temperatures
ranging from 1050 to 1060 $^\circ$C. Finally, a prolonged anneal
in flowing O$\ped{2}$ at 1060 $^\circ$C was performed, during
which the material densifies, granularity is reduced and ordering
within the crystal structure develops. For further details on the
preparation technique see Ref.\cite{Masini}.

\begin{figure}[ht]
\begin{center}
\includegraphics[keepaspectratio, width=0.5\textwidth]{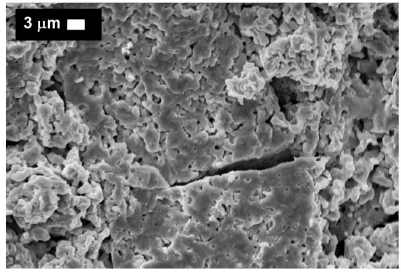}
\end{center}
\vspace{-3mm} \caption{SEM image of a polycristalline
RuSr$\ped{2}$GdCu$\ped{2}$O$\ped{8}$ sample similar to that used
for our measurements, and synthesized in Genoa (from
Ref.~\cite{Masini}).}\label{SEMruteno}
\vspace{-3mm}
\end{figure}

Figure~\ref{SEMruteno} reports a SEM image of one of the samples
studied. It is clearly seen that the thermal
treatment~\cite{Masini} ensures a partial sinterization (some
grains begins to coalesce in big aggregates) but a matrix of
weakly connected grains is still present. This kind of aggregation
can be also perceived by optical inspections with AFM.

\begin{figure}[!ht]
\begin{center}
\includegraphics[keepaspectratio, width=0.45\textwidth]{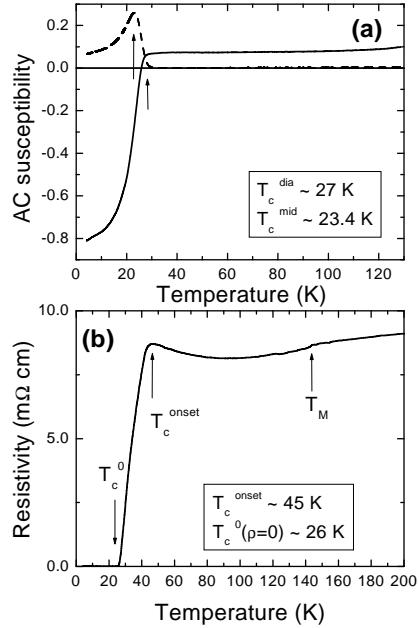}
\vspace{-3mm} \caption{(a) An example of AC suseptibility
transition of a RuSr$\ped{2}$GdCu$\ped{2}$O$\ped{8}$ sample. The
values of the relevant critical temperatures are also reported.
(b) Resistivity vs. temperature for one of our samples. The
relevant temperatures corresponding to the magnetic and
superconducting transitions are also indicated.}\label{suscres}
\end{center}
\vspace{-3mm}
\end{figure}

Previous to point-contact fabrication and PCS measurements, we
quickly characterized the samples from the superconducting and
electrical point of view, by means of resistivity and AC
susceptibility measurements. Figure~\ref{suscres}a reports the
temperature dependence of the real and imaginary parts of the
susceptibility, measured in the absence of DC magnetic field. The
measurements were carried out in a home-made susceptometer
designed to allow measurements also in very small samples (sub-mm
size). The temperature is given by a SMD-type resistor mounted in
the close proximity of the coil containing the sample.
Unfortunately, it is calibrated only up to 130~K so that the peak
in $\chi'$ at about 130~K is only partly visible. The correction
for the demagnetizing factor and the correct volume are taken into
account in the conversion from the imbalance voltage to the
susceptibility. It is clear that, in this kind of measurements,
what we actually observe is the volume fraction that is
diamagnetically shielded, and \emph{not} the superconducting
fraction. In this sense, we can safely say that in our samples the
superconducting volume is \emph{at most} 80\% of the total volume
(see Figure~\ref{suscres}a). The diamagnetic transition sets on at
$T\ped{c} ^{dia}\simeq 27$~K, while the middle point
$T\ped{c}^{mid}$ (the maximum peak in the imaginary part curves)
is estimated at $T\ped{c}^{mid}\simeq 23.4$~K.

The resistivity measurements were carried out by using the
conventional four-probe technique and by using AC current to
cancel out thermoelectric effects. The voltage signal was recorded
by using a EG\&G lock-in amplifier that allows using very small
frequencies. In particular, the $\rho(T)$ curve reported in
Figure~\ref{suscres}b was obtained with a very small current
intensity (3 mA p-p) to reduce possible heating effects, with a
frequency of 3.0~Hz. The curve witnesses the good quality of the
samples. It is well known that the physical properties of this
rutheno-cuprate material -- and in particular its critical
temperature -- are strongly dependent on the details of the
preparation procedure. In the present case, the temperature at
which the resistance goes to zero ($T\ped{c}^{0}=26$~K) is rather
high if compared to other data from literature. The transition
sets on at $T\ped{c}^{onset}=45$~K, so that $\Delta T\ped{c}=19$~K
is the transition width. The change in slope usually attributed to
the onset of the magnetic order at $T\ped{M}$ is also shown.

\subsection{Point-contact measurements}
Due to the granularity of the material and its only partial
sinterization, the surface of the samples looked rather rough and
brittle. As a result, point-contact measurements turned out to be
rather difficult. We tried with tips of different materials (Au,
Pt, Pt-Ir and so on), made by electrochemical etching (with a
25\%HNO$_3$+ 75\%HCl solution) starting from wires of different
thickness. We also tried to use, instead of the tip, Ag paste or
In flakes to make the contacts. In the end, we achieved the best
reproducibility of the conductance curves in point contacts
obtained by using tips (of either Au or Pt-Ir) made by starting
from thick wires ($\varnothing \leq 0.1$~mm), that could apply a
rather large pressure on the sample. In these contacts, we were
able to follow the temperature evolution of the conductance
curves, from 4.2 K up to the critical temperature of the junction.

Once established the contact between tip and sample, we measured
the DC $I-V$ characteristic of the contact and then numerically
calculated the differential conductance, i.e. the first derivative
d$I$/d$V$ as a function of the voltage bias $V$. To allow the
comparison to theoretical models, the resulting curves were then
normalized to the normal-state conductance.
\begin{figure}[ht]
\begin{center}
\includegraphics[keepaspectratio, width=0.45\textwidth]{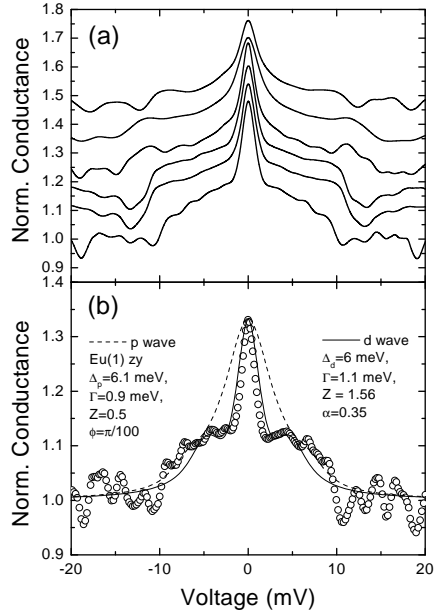}
\end{center}
\vspace{-3mm} \caption{(a) normalized conductance curves obtained
by PCS with Pt-Ir tips at low temperature. (b) an example of
normalized conductance curve fitted with a pure $d$-wave symmetry
or with a pure $p$-wave symmetry. The best fit was due to the pure
$d$-wave order parameter symmetry.}\label{LowT} \vspace{-3mm}
\end{figure}
Figure~\ref{LowT}a reports some examples of normalized conductance
curves measured at low temperature ($T=4.2$~K). The curves are
vertically shifted for clarity. All of them feature a zero-bias
conductance peak (ZBCP), and more or less pronounced shoulders at
finite voltage. The zero-bias peak is very similar to that
produced by Andreev reflection in normal metal/$d$-wave
superconductor junctions, when the positive interference between
incoming and reflected electron gives rise to the zero-energy
Andreev bound states. Therefore, the presence of a ZBCP is
suggestive of a $d$-wave order parameter in Ru-1212.

To investigate in more detail this possibility, we tried to fit
the low-temperature normalized conductance curves by using the
Blonder-Tinkham-Klapwijk model \cite{BTK} generalized to
non-conventional order parameter symmetries \cite{Tanaka}. For the
time being, we only considered the simplest possible symmetries:
$d$-wave (because of the analogy with other cuprates) and $p$-wave
(suggested by the comparison with the superconducting ruthenate
Sr$_2$RuO$_4$). In Figure~\ref{LowT}b an experimental conductance
curve (symbols) is reported together with the best-fitting curves
obtained with pure $d$-wave (solid line) and pure $p$-wave (dashed
line) order parameter symmetry. The values of the best-fitting
parameters are also reported in the legend.

In the case of the $d$-wave fit, the free parameters are: the
maximum gap amplitude $\Delta$, the barrier transparency parameter
$Z$ (which is proportional to the potential barrier height at the
interface) and the angle $\alpha$ the direction of current
injection makes with the normal to the interface. In addition to
these parameters, a phenomenological broadening parameter $\Gamma$
was needed, that mimics pair-breaking effects due, for example, to
non-magnetic impurities. Notice that, however, $\Gamma$ is much
smaller than the gap amplitude ($\Delta\ped{d}=6.0$~meV) given by
the fit.

As far as the $p$-wave symmetry is concerned, we chose one of the
triplet $p$-wave pair potential symmetries (the so-called
$E\ped{u}$ states) proposed by Machida \emph{et al} \cite{Machida}
and by Sigrist and Zhitomirsky \cite{Sigrist}. The free fitting
parameters are: the gap amplitude $\Delta$, the barrier parameter
$Z$, the azimuthal angle $\phi$ and, as in the previous case, the
broadening parameter $\Gamma$. Interestingly, the amplitude of the
$p$-wave gap obtained from the fit ($\Delta\ped{p}=6.1$~meV) is
consistent with that obtained with the $d$-wave symmetry.

It is clearly seen in Figure~\ref{LowT}b that the $d$-wave fit
works much better than the $p$-wave one. In particular, the
$d$-wave curve accounts for both the height and the width of the
zero-bias conductance peak, and also follows rather well the
shoulders at finite energy. On the contrary, the $p$-wave curve
has a smoother bell-shaped behaviour that cannot reproduce the
main features of the experimental conductance curve. Thus, the
$d$-wave symmetry is much more compatible with our findings. This
fact indicates that the nature of superconductivity in the Ru-1212
compound is by far more similar to other cuprates than to
ruthenates. In fact: i) the $d$-wave symmetry is common among
cuprates, for example in BSCCO; ii) if the order parameter is
$d$-wave, the Cooper pairs turn out to be spin singlets as in
conventional superconductors and HTSC, and not exotic spin-triplet
states as in Sr$_2$RuO$_4$. Indirectly, this similarity with other
cuprates supports the evidences of a confinement of
superconductivity on the CuO$_2$ layers (that are common to
Ru-1212 and copper oxides), with no contribution from the Ru-O
sublattice \cite{Pozek}.
\begin{figure}[ht]
\begin{center}
\includegraphics[keepaspectratio, width=0.45\textwidth]{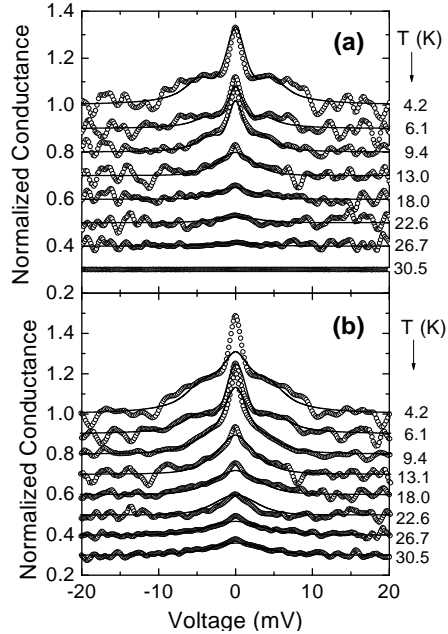}
\vspace{-3mm} \caption{Temperature dependence of the conductance
curves of point-contacts obtained by using a Pt-Ir and a Au tip,
respectively. The diameter of the starting wires in which the tips
were made by electrochemical process was $\varnothing=0.1$ and
$\varnothing=0.125$~mm, respectively. The temperature
corresponding to each curve is reported on the right side of the
panel. The experimental conductance curves (symbols) are compared
to the theoretical best-fitting BTK curves, calculated for a
$d$-wave order parameter.}\label{TempDep}
\end{center}
\vspace{-3mm}
\end{figure}

When the junction was sufficiently stable, we could follow the
evolution of its conductance curves on heating, from 4.2~K up to
the critical temperature of the junction, where the
Andreev-reflection features disappear. Figure~\ref{TempDep}
reports two examples of temperature dependence of the normalized
conductance curves. The curves are vertically shifted for clarity
and ordered for increasing values of the temperature. Superimposed
to the experimental data (symbols) we also show the best-fitting
$d$-wave symmetry curves. It is clear that, in general, the fit is
satisfactory. In these few cases where the agreement between
theoretical and experimental data is poor (see, for example, the
upper curve in panel (b)) the choice had to be made between the
theoretical curve that best fits the ZBCP and the one that,
instead, follows the shoulders.

\begin{figure}[ht]
\begin{center}
\includegraphics[keepaspectratio, width=0.5\textwidth]{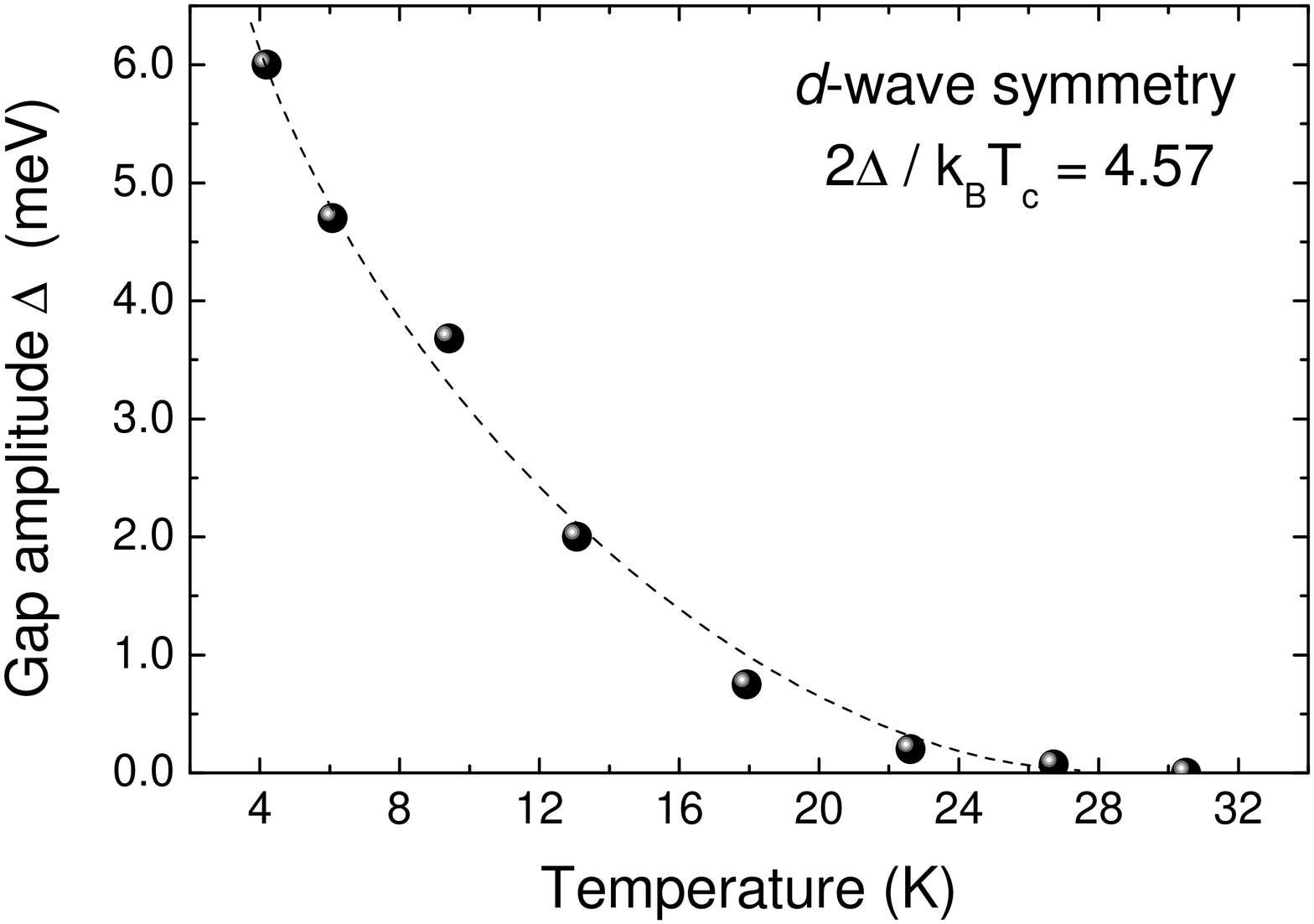}
\vspace{-3mm} \caption{Superconducting gap values as a  function
of the temperature, as obtained from the generalized BTK fit of
the experimental data in Figure~\ref{TempDep}a. The line is only a
guide to the eye.}\label{Gap}
\end{center}
\vspace{-3mm}
\end{figure}

It is clearly seen that, in both cases, the thermal evolution of
the conductance curves is approximately the same, with a
progressive decrease of the finite-voltage shoulders and of the
zero-bias peak. As a matter of fact, also the behaviour of the
best-fit parameters as a function of the temperature is very
similar in the two cases. First of all, the broadening parameter
$\Gamma$ decreases quickly and goes to zero well before the
critical temperature. Second, the gap amplitude starts decreasing
at low temperature, contrary to what expected in the conventional
BCS theory, and goes to zero at $T\ped{c}$ in a sub-linear way.
Figure~\ref{Gap} reports the complete $\Delta(T)$ curve for the
conductances of Figure~\ref{TempDep}a, that are better fitted by
the model. In the other case, the behaviour is similar but the gap
values are slightly greater at intermediate temperatures.

\section{Conclusions}
In conclusion, we have presented some preliminary results of
point-contact spectroscopy in the hybrid ruthenate-cuprate
superconductor RuSr$_2$GdCu$_2$O$_8$. The low temperature (4.2 K)
conductance curves present a clear zero-bias conductance peak,
that decreases and goes to zero at the increase of the
temperature. The experimental conductance curves turn out to be
compatible with a pure $d$-wave symmetry of the order parameter,
which suggests a tighter similarity of this compound to the other
cuprates rather than to the ruthenate Sr$_2$RuO$_4$. It is worth
reminding that, however, alternative explanations cannot be ruled
out, due to the complexity of the material under study and in
particular to its magnetic properties.

One of the authors (V.A.S.) acknowledges the support from RFBR
(project N. 02-02-17133) and the Ministry of Science and
Technologies of the Russian Federation (contract N.
40.012.1.1.1357).

\end{document}